\documentclass[prl,amsmath,superscriptaddress,amssymb,twocolumn,floatfix,aps,showpacs,showkeys]{revtex4}

\usepackage{amssymb} 
\usepackage{amssymb}
\usepackage{epsfig}
\usepackage{graphicx} % or graphicx 

\def \figwidth{8.5cm}

\begin{document}

\title{Model independent extraction of the pole and Breit-Wigner resonance parameters}
\author{S. Ceci}
%\affiliation{Rudjer Bo\v{s}kovi\'{c} Institute, Bijeni\v{c}ka  54, HR-10000 Zagreb, Croatia}
\author{M. Korolija}
%\affiliation{Rudjer Bo\v{s}kovi\'{c} Institute, Bijeni\v{c}ka  54, HR-10000 Zagreb, Croatia}
\author{B. Zauner}
\affiliation{Rudjer Bo\v{s}kovi\'{c} Institute, Bijeni\v{c}ka  54, HR-10000 Zagreb, Croatia}

\date{\today}                                           % Activate to display a given date or no date

\begin{abstract}
We show that a slightly modified Breit-Wigner formula can successfully describe the total cross section even for the broad resonances, from light $\rho(770)$ to the heavy Z boson. In addition to mass, width, and branching fraction, we include another resonance parameter that turns out to be directly related to the pole residue phase. The new formula has two mathematically equivalent forms: one with the pole, and the other with the Breit-Wigner parameters. 
\end{abstract}
%\begin{keyword}
% Scattering matrix\sep  Resonances\sep Heavy quarkonia\sep Z bosons.
%\PACS 
   %%11.55.Bq \sep	%Analytic properties of S matrix 
%%  11.80.Gw, 	%Multichannel scattering
   %% 12.40.Yx\sep 	%Hadron mass models and calculations 
%% 13.25.Gv \sep	%Decays of J/psi, and other quarkonia 
%%  13.75.Gx, 	%Pion-baryon interactions
   %% 14.20.Gk\sep	%Baryon resonances with <i>S</i>=0
%% 14.40.Be, 	%Light mesons (S=C=B=0) 
%% 14.40.Pq\sep 	%Heavy quarkonia 
   %% 14.70.Hp. 	%Z bosons 
%\end{keyword}

\maketitle

%\section{Resonances}
Resonances are unstable particles usually observed as bell-shaped structures in scattering cross sections of their decay products. For a simple narrow resonance, its fundamental properties correspond to the visible cross-section features: mass $M$ is at the peak position, and decay width $\Gamma$ is the width of the bell shape. These parameters, along with the branching fraction $x$, are known as the Breit-Wigner parameters \cite{BW}. In reality, resonance peaks may be very broad, and the shape so deformed that it is not at all clear where exactly is the mass, or what would be the width of that resonance. In such cases resonance parameters are treated as energy dependent functions. These functions are often defined differently for different resonances. For example, in the case of $\rho(770)$ resonance in the $\pi \pi$ channel, modern analyses include the pion-pion P-wave potential barrier (momentum to three halves) in the energy dependent width \cite{GS}, while in the case of $Z$ boson the width function is 
proportional to the energy squared \cite{Sir91}. 

With such model dependent parameterizations, the simple connection between physical properties of a resonance and its model parameters is lost, and the choice of the "proper" resonance parameters becomes the matter of preference. There are many definitions for Breit-Wigner mass, which is assumed by some to be the proper resonance physical property. Other will prefer the real part of pole position in the complex energy plane. Some will even define the resonance mass to be something unrelated to these two most common definitions, as we will soon see, or assume that there is no difference between poles and Breit-Wigner parameters whatsoever. All that makes the comparison between cited resonance parameters quite confusing and potentially hinders the direct comparison between microscopic theoretical predictions (such is \cite{Dur08}) and experimentally obtained resonance properties \cite{PDG}. 

To clarify this situation we try to devise a simple model-independent formula for resonant scattering, with well defined resonance physical properties, which will be capable of successfully fitting the realistic data for broad resonances.

In this letter we show how to dramatically improve the simple Breit-Wigner formula by incorporating in it just one additional (phase) parameter. This new formula has two equivalent forms that can be used to estimate either pole or Breit-Wigner parameters in a model independent way. 

We begin our analysis by noting that resonant cross section is commonly parameterized by a simple Breit-Wigner formula \cite{BW}
\begin{equation}
\sigma =\frac{4\pi}{q^2}\frac{2J+1}{(2s_1+1)(2s_2+1)}\, |A|^2,
\end{equation}
where $q$ is a c.m.~momentum, $J$ is the spin of the resonance, while $s_1$ and $s_2$ are spins of the two incoming particles. Resonant scattering amplitude $A$ is given by
\begin{equation}\label{eq:PDGparameterization}
A=\frac{x\, \Gamma/2}{M-W-i\,\Gamma/2},
\end{equation}
where $M$ is the resonant mass, $\Gamma$ is the total decay width, $x$ is the branching fraction to a particular channel (for inelastic scattering it is $\sqrt{x_{\mathrm{in}}\,x_{\mathrm{out}}}$), and W is c.m.~energy. 

This simple parameterization cannot describe most of the realistic cross sections since the resonance shapes are seldom symmetric. To fix this, a background contribution is usually added. Unfortunately, there is no standard way to add background, but polynomials in $W^2$ (i.e., Mandelstam $s$) are commonly used in the literature (see e.g.~\cite{BES06}) 
\begin{equation}\label{eq:polybg}
|A|^2\rightarrow|A|^2+\sum_{k=0}^n B_k\,W^{2k}.
\end{equation}

To extract the resonance parameters, we do local fits (in energy) of this parameterization to a broad range of data points in the vicinity of the resonance peak. To estimate the proper order $n$ of polynomial background, we vary endpoints of the data range and check the convergence of the physical fit parameters: $M$, $\Gamma$, and $x$. Goodness of the convergence is estimated by calculating $c_{n,l}$ parameters for each data range and for all polynomial orders $n$ and $l$ 
\begin{equation}\label{eq:crit1}
c_{n,l}=\sum_{y=m,\Gamma,x}(y_l-y_n)^2/y_n^2.
\end{equation}
Smaller $c_{n,l}$ means better convergence. 

To avoid false positive convergence signals as much as possible, we demand good convergence not just for two, but for three consecutive polynomial orders by using 
\begin{equation}\label{eq:crit2}
c_{n}=c_{n,n+1}+c_{n,n+2}.
\end{equation}
Final result is the one having the smallest reduced $\chi^2_R$ among several fits (we use ten) with lowest convergence parameters $c_n$. When statistical errors turn out to be unrealistically small due to the dataset issues, the spread in extracted pole parameter values is used to estimate parameters errors. 

To test this extraction approach, we analyze five broad resonances with well known properties, and masses ranging from less then 1 GeV, to almost 100 GeV. For $\Delta(1232)$ and $N(1440)$, we analyze GWU \cite{GWU} $\pi N$ elastic partial-wave amplitudes squared. For $\rho(770)$ and $Z$ boson we analyze $e^+e^-$ scattering ratio R (between hadronic an muonic channels) from PDG compilation \cite{PDG}, and for $\Upsilon(11020)$  the new {\sc BaBar} data \cite{Aub09}. %Results are shown in Table \ref{tab:PDGnotgood}.

%\section{Energy dependent parameters}
%As expected, 
Using the Breit-Wigner parameterization (\ref{eq:PDGparameterization}) on broad resonances does not produce very good results. Therefore, in the advanced approaches, resonance width $\Gamma$ (and other parameters) are considered to be energy dependent, which drastically improves the fit. However, parameterization then becomes model dependent, obfuscating the connection between the model parameters and physical properties of the resonance. We want to find a simple model independent form, as close to the original Breit-Wigner parameterization as possible, that will be capable of successfully fitting the realistic data for broad resonances. To do so, we assume that the numerator and the denominator in Rel.~(\ref{eq:PDGparameterization}) are functions of energy, expand them and keep only the linear terms. The amplitude $A$ becomes
\begin{equation}\label{eq:res.amplitude}
A = \frac{x_p\,\Gamma_p/2\, \,e^{i\theta_p}}{M_p-W-i\,\Gamma_p/2}+|A_{B}|\,e^{i\theta_{B}},
\end{equation}
which turns out to be the lowest order Laurent expansion of amplitude A about its pole position at \mbox{$W= M_p-i\,\Gamma_p/2$}. Therefore, $M_p$ and $\Gamma_p$ are pole mass and width, while $x_p\,\Gamma_p/2$ and $\theta_p$ are the complex residue magnitude and phase, respectively. (Note that we use the standard convention for residue phase $\theta_p$, as used in PDG \cite{PDG}, which differs from the mathematical residue phase by $\pm\pi$.) Three additional fit parameters are residue phase $\theta_p$, (coherently added) background magnitude $|A_B|$, and the background phase $\theta_B$. We can extract only the relative phase $\delta_{p}=\theta_{p}-\theta_{B}$, since absolute square of this amplitude will be compared to the data. In order to ease the numerical analysis, we rewrite the new parameterization in a compact form
\begin{equation}\label{eq:betterparameterization}
|A|^2 = |A_B|^2\frac{(\mu-W)^2+\lambda^2}{(M_p-W)^2+\Gamma_p^2/4},
\end{equation}
where $\mu$ and $\lambda$ are simple fit parameters related to the pole parameters through 
\begin{align}
x_p\,\sin\delta_p&=|A_B|\,\frac{\Gamma_p/2-|\lambda|}{\Gamma_p/2},\\
x_p\,\cos\delta_p&=-|A_B|\,\frac{M_p-\mu}{\Gamma_p/2}.
%\tan(\theta-\theta_{bg})&=-\frac{\Gamma/2-\lambda}{m-\mu},\\
%|r|&=\beta\,\sqrt{(m-\mu)^2+(\Gamma/2-\lambda)^2}.
\end{align}

Using parameterization (\ref{eq:betterparameterization}), we should be able to extract pole mass, width, branching fraction, magnitude of the background amplitude, and the relative phase from the data. We again use the same polynomial background from relation (\ref{eq:polybg}) and convergence criteria from  relations (\ref{eq:crit1}) and (\ref{eq:crit2}). However, at the very beginning of this analysis we stumbled upon a problem with our fits. When we fitted $\Delta(1232)$ resonance, parameter $\lambda$ was rather unstable, ranging from zero to several thousands MeV. In addition, fits often did not converge, even for a carefully chosen initial values. 

We looked into it more closely and realised that since $\Delta(1232)$ is almost elastic resonance (decaying by more than 99 percent to $\pi N$ channel), $\lambda$ should be zero due to the elastic two-body unitarity condition \mbox{$A^\dag A=\mathrm{Im}\,A$}. When $\lambda$ was set to zero, everything worked almost perfectly. It is important to note  that $x_p$ should be 1 for elastic resonances, again due to the unitarity, but setting $\lambda$ to zero does not imply that $x_p$ is 1.

Things became really interesting when we tried to extract Z boson parameters from $e^+e^-$ scattering data. The unstable $\lambda$ behavior seen in the case of $\Delta(1232)$ was observed again, even though Z boson is definitely not an elastic resonance. Fit could not be stabilized, and eventually we tried $\lambda=0$ again ($x_p$ still can take care of inelasticity). This choice smoothed the fitting procedure, and the extracted resonance parameters were in excellent agreement with the PDG (pole) estimates \cite{PDG}. Assuming that $\lambda=0$ for other processes, we rewrite amplitude defined in Eq.~(\ref{eq:res.amplitude}) as:
 \begin{equation}\label{eq:newamplitude}
A = x_p\,e^{i\eta}\left(\frac{\Gamma_p/2 \,\, e^{2i\delta_p}} {M_p-W-i\,\Gamma_p/2} + e^{i\delta_p}\sin\delta_p\right),
\end{equation}
with unmeasurable overall phase $\eta$ equal to $2\theta_{B}-\theta_p$. Square of this amplitude is then
 \begin{equation}\label{eq:pole}
 |A|^2 = x_p^2\,\frac{\left[(M_p-W)\,\sin\delta_p+\Gamma_p/2\,\cos\delta_p\right]^2}{(M_p-W)^2+\Gamma_p^2/4}.
\end{equation}

\begin{figure}[h!]
\includegraphics[width=\figwidth]{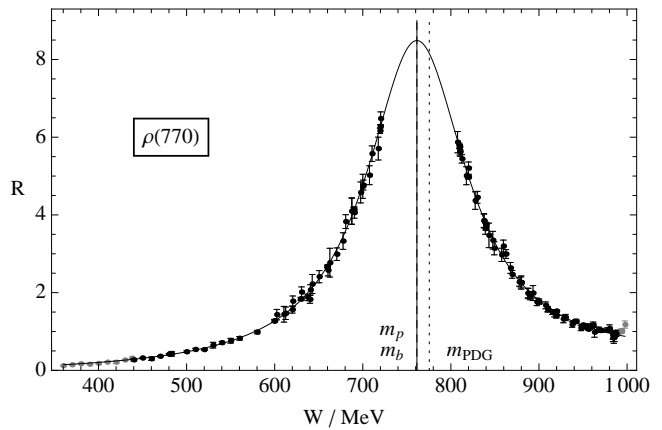}
\caption{Fitting pole parameterization (\ref{eq:pole}) to the data \cite{PDG}. The pole and BW masses have almost the same value, quite different from PDG estimate (dotted line). (We removed the data from the peak to eliminate the influence of the $\omega(782)$ resonance.) \label{fig:results5}}
\end{figure}

\begin{figure}[h!]
\includegraphics[width=\figwidth]{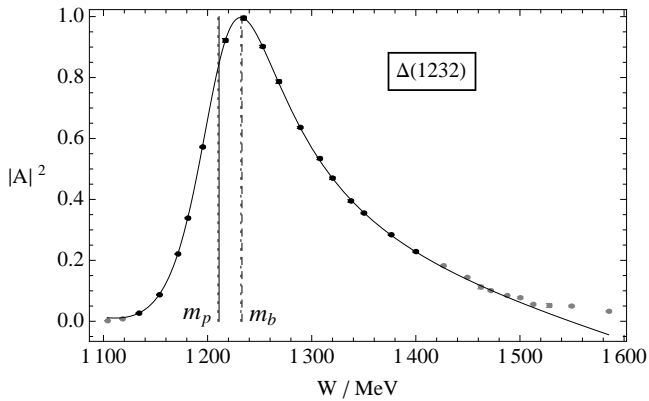}
\caption{Fitting pole parameterization (\ref{eq:pole}) to the SAID data \cite{GWU}. Pole (solid) and BW mass (dashed) are clearly distinct, while the PDG estimates (dotted lines) are indistinguishable from them. All the data in this figure are analyzed, but only the black data points are used in the best fit. \label{fig:results1}}
\end{figure}

\begin{figure}[h!]
\includegraphics[width=\figwidth]{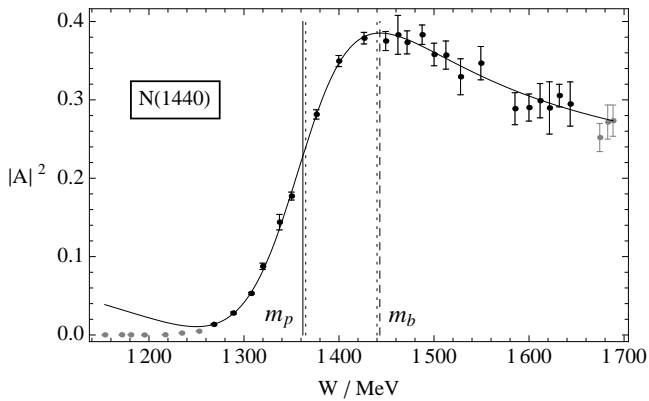}
\caption{Fitting pole parameterization (\ref{eq:pole}) to the SAID data \cite{GWU}. This resonance has the largest difference between pole and BW mass. PDG estimates (dotted liness) are  consistent with pole (solid) and BW (dashed) parameters. \label{fig:results2}}

\end{figure}

\begin{figure}[h!]
\includegraphics[width=\figwidth]{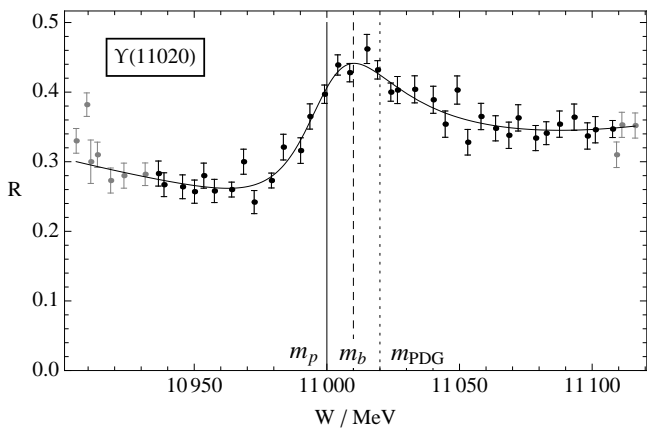}
\caption{Fitting pole parameterization (\ref{eq:pole}) to the {\sc BaBar}  data \cite{Aub09}. Pole and BW masses are clearly distinct, but PDG estimate coincides with neither of them. \label{fig:results14}}
\end{figure}

\begin{figure}[h!]
\includegraphics[width=\figwidth]{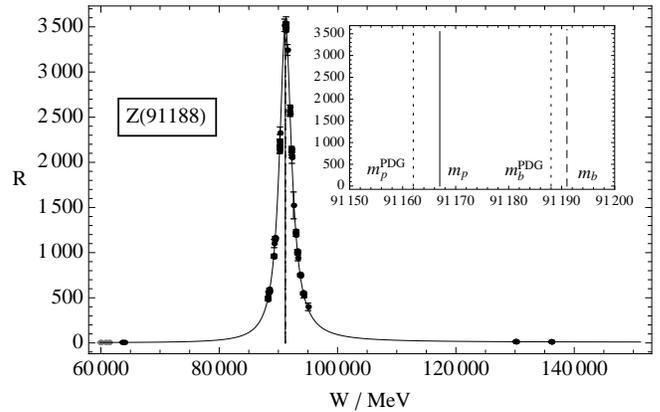}
\caption{Fitting pole parameterization (\ref{eq:pole}) to the data \cite{PDG}. The pole and BW parameters are consistent with their PDG estimates (dotted lines).  \label{fig:results11}}
\end{figure}

We know that for $\Delta(1232)$, the overall phase $\eta$ iz zero due to unitarity, which means that $\theta_{B} = \delta_p$ and $\theta_p=2\delta_p$. We  compare our results for $2\delta_p$ to published results of $\theta_p$ for other analyzed resonances to check whether the same relation is valid for them as well.  Roper resonance N(1440) is $\pi N$ resonance with $\pi N$ branching fraction $x$ estimated to 65\%, and the $2\delta_p$ value of $-81^\circ$ is a surprisingly close to the newest residue phase estimate  $-85^\circ$ from \cite{PDG}. For the Z boson, these two values are even closer: $2\delta_p$ is $-2.2^\circ$, while $\theta_p$ is $-2.35^\circ$. Extracted masses and widths are much closer to the pole parameters listed in the literature, than to the Breit-Wigner ones, as can be seen in Table \ref{tab:poles}. The best fits for all analyzed resonances are shown in Figures \ref{fig:results5}-\ref{fig:results11}.

\begin{table}[h!]
\caption{\label{tab:poles} Resonance pole parameters extracted by using pole formula (\ref{eq:pole}). Our $2\delta_p$ is compared to the residue phase $\theta_p$ from the literature. PDG pole estimates are from Ref.~\cite{PDG}. The $\rho$ meson pole and the Z boson residue phase are estimated by analytic continuation of Gounaris-Sakurai \cite{GS} and Breit-Wigner \cite{PDG} parameterization, respectively. }
\begin{ruledtabular}
\begin{tabular}{lllll}
Resonance & $M_p$ / MeV & $\Gamma_p$ / MeV & $x_p$ / \% & $2\delta_p$ / $^o$\\\hline
$\rho(770)$                &  762   $\pm$  1   &   138  $\pm$  2   &   0.71    &   1  $\pm$  1   \\
%PDG                               &  775.5   $\pm$  0.3  &  146.2   $\pm$  0.7   &   N/A   &  N/A   \\
\scriptsize POLE                               &  763   &   144    & \scriptsize  N/A   & \scriptsize N/A   \\\hline 
%$\Delta(1232) / \sigma$ &     $\pm$     &     $\pm$     &     $\pm$     &     $\pm$     \\
$\Delta(1232)$               &  1211   $\pm$ 1    &  102  $\pm$  1   &  103   $\pm$  1   &  $-$47   $\pm$ 1   \\
\scriptsize PDG/POLE                       &  1210   $\pm$  1   &  100   $\pm$  2   &  104 $\pm$ 2 &   $-$47  $\pm$  1   \\\hline
$N(1440) $           &  1362   $\pm$  5   &   191  $\pm$  10   &  61   $\pm$  4   &  $-$81  $\pm$  10   \\
\scriptsize PDG/POLE               &  1365   $\pm$  15   &  190   $\pm$  30   &  65   $\pm$  10   &  $-$85   $\pm$  $^{15}_{10}$   \\\hline%\\\\
$\Upsilon(11020)$         &   11000  $\pm$  2   &  43   $\pm$  6   &  0.10   &  $-$52   $\pm$  8   \\
%PDG                               &   11019  $\pm$  8   &  79   $\pm$  16   &  N/A  & N/A  \\
 {\scriptsize\sc BaBar}  \cite{Aub09}                   &   10996  $\pm$  2   &  37   $\pm$  3     & \scriptsize N/A  & \scriptsize N/A  \\\hline

$Z(91188)$                    &  91167   $\pm$  6   &  2493   $\pm$  5   &  15.4    &   $-$2.2  $\pm$   0.2  \\
%PDG                               &  91188   $\pm$  2   &  2495   $\pm$  2   &  N/A   &     N/A     \\
\scriptsize PDG/POLE                               &  91162   $\pm$  2   &  2494   $\pm$  2   &  \scriptsize N/A   &     $-2.35$    \\
\end{tabular}
\end{ruledtabular}
\end{table}

The new pole parameterization (\ref{eq:pole}) may be used instead of (\ref{eq:PDGparameterization}), since it works much better and adds only one quite important parameter $\delta_p$. This parameter is the main ingredient of the shape of the resonance contribution to the cross section. When $\delta_p$ is equal to zero, the new pole and the old simple Breit-Wigner parameterization are exactly the same.

%\section{Breit-Wigner parameterization}
We began this study in the first place to find an improved Breit-Wigner parameterization, but ended up with pole parameterization instead. Following the original notion of Breit and Wigner, that the resonance mass is at the peak position \cite{BW}, and by noting the convenient form of our pole parameterization (\ref{eq:newamplitude}) that looks very similar to a single channel elastic amplitude (apart from $x_p\neq1$ and $\eta\neq0$). We now define the new Breit-Wigner parameters as single-channel K-matrix pole $M_b$, residue $\Gamma_b$, branching fraction $x_b$, and background phase $\delta_b$
\begin{align}
K&= \frac{\Gamma_{b}/2}{M_{b}-W} + \tan\delta_{b},\\
A&=x_{b}\,\frac{K}{1-iK},\\
|A|^2 &=x_{b}^2\frac{\left[\Gamma_{b}/2+(M_{b}-W)\tan{\delta_{b}}\right]^2}{(M_{b}-W)^2+\left[\Gamma_{b}/2+(M_{b}-W)\tan{\delta_{b}}\right]^2}. \label{eq:BW}
\end{align}

In this form, $x_{b}$ and $\delta_{b}$ will be mathematically equal to $x_p$ and $\delta_p$, respectively. When we fit parameterization (\ref{eq:BW}) to the data, extracted fit parameters $M_{b}$, $\Gamma_{b}$, and $x_{b}$ are consistent with Breit-Wigner parameters in PDG \cite{PDG}, as is clearly visible from Table \ref{tab:breitwigner}. As expected,  $x_{b}$ and $\delta_{b}$ have almost exactly equal values as their pole counterparts $x_p$ and $\delta_p$ in Table \ref{tab:poles}. 
Furthermore, the extracted pole and Breit-Wigner parameters are interrelated through Manley relations \cite{Man95}
\begin{align}
M_{b}&=M_p-\Gamma_p/2\,\,\tan\delta_p,\\
\Gamma_{b}&=\Gamma_p/\cos^2\delta_p. 
\end{align}

\begin{table}[h!]
\caption{\label{tab:breitwigner}  Resonance parameters extracted by using new Breit-Wigner formula (\ref{eq:BW}). PDG estimates are from Ref.~\cite{PDG}.}
\begin{ruledtabular}
\begin{tabular}{lllll}
Resonance                    & $M_{b}$ / MeV         & $\Gamma_{b}$ / MeV & $x_{b}$ / \% & $2\delta_{b}$ / $^o$ \\\hline
$\rho(770)$                &  761   $\pm$  1  &    139  $\pm$  2   &   0.71    &   0  $\pm$ 1   \\
\scriptsize  PDG                               &   \scriptsize 775.5   $\pm$  0.3  &   \scriptsize 146.2   $\pm$  0.7   & 0.69   & \scriptsize N/A   \\
\hline
$\Delta(1232)$               &   1233  $\pm$    1         &   120  $\pm$   1  &  102   $\pm$   1  &  $-$46  $\pm$ 1    \\
\scriptsize  PDG/BW                         &  1232   $\pm$  2   &  117   $\pm$  3   &  100   &  \scriptsize N/A   \\\hline
$N(1440) $                    &  1443 $\pm$ 2   &  325 $\pm$ 11  &  61 $\pm$ 4   &   $-$80 $\pm$ 2  \\
\scriptsize  PDG/BW                        &  1440   $\pm$  $^{30}_{20}$   &  300   $\pm$  $^{150}_{100}$   &  65   $\pm$  10   & \scriptsize N/A   \\\hline
$\Upsilon(11020)$           &       11010        $\pm$    2    &      53        $\pm$  8   &    0.10  &  $-$52 $\pm$ 8     \\
\scriptsize  PDG                                &   11019  $\pm$  8    &  79  $\pm$  16   &   \scriptsize N/A   &  \scriptsize  N/A      \\\hline
$Z(91188)$                     & 91191 $\pm$  5   &   2494           $\pm$ 5    &   15.4  &  $-$2.2 $\pm$ 0.2     \\
\scriptsize  PDG/BW                         &  91188  $\pm$  2   &  2495   $\pm$  2   & 15.3   &  \scriptsize   N/A       
\end{tabular}
\end{ruledtabular}
\end{table}


These Breit-Wigner parameters are uniquely defined and model independent, with directly observable mass as the peak of the squared amplitude $|A|^2$. However, they strongly depend on phase $\delta_p$, which may change from reaction to reaction. That means that for the same pole position, there will be different Breit-Wigner masses and widths in different channels. Therefore, it is more practical to have one (pole) mass and width for each resonance in particle data tables \cite{PDG} than stockpiling different (Breit-Wigner) masses and widths for each process in which the resonance contributes.

In our study, there are two resonances that show systematic discrepancy between the PDG mass estimates \cite{PDG} and our results presented here: the $\rho(770)$ and the $\Upsilon(11020)$. For $\rho$ meson an alternative parameterization by Gounaris and Sakurai  \cite{GS} is used, where mass and width are defined somewhat unconventionally to take into account its mixing with $\omega(782)$ and $\rho(1450)$. $\Upsilon(11020)$ mass and width were fit parameters to a Gaussian with a relativistic tail \cite{CUSBCLEO}. For both resonances cited values in the PDG tables are neither masses consistent with the original Breit-Wigner idea, being at the peak of the resonance, nor the pole positions. Since the resonance parameters are collected in PDG tables to be used as an input for various models and for comparison between theory and experiment, placing all these resonance parameters in a single table may generally create considerable confusion. This confusion is evident in the $\rho$ meson case where in the table 
with predominantly Gounaris-Sakurai masses (about 775 MeV) one can find pole masses (roughly 760 MeV).  

In conclusion, we have shown here that the original Breit-Wigner formula may be drastically improved by including a single additional (phase) parameter $\delta_p$. Our results suggest that parameter $\delta_p$ seems to be equal to the half of the resonance residue phase $\theta_p$, regardless of the resonance inelasticity. This new formula has two equivalent forms that can be used to estimate either the pole, or the Breit-Wigner parameters in a model independent way. Having both forms enabled us to learn that in the PDG tables \cite{PDG} there are values that do not correspond neither to pole, nor to Breit-Wigner parameters. Such an outcome undermines the proper matching between microscopic theories (e.g., lattice QCD \cite{Dur08}) and experiment.

\end{document}